%% file: 2017-wdt.tex
\documentclass[times, 3p, twocolumn]{elsarticle}
\usepackage{tabularx}
\usepackage{subcaption}
\usepackage{todonotes}
\usepackage{pifont}
\usepackage{natbib}
\usepackage{fleqn}
\usepackage{booktabs}
\usepackage{graphicx}
\usepackage{algorithm}
\usepackage{algpseudocode}
\usepackage{tikz}
\usepackage{amsmath}
\usepackage{amssymb}
\usepackage{bm}
\usepackage{xcolor}
\definecolor{boxgray}{HTML}{F0F0F0}
\usepackage{mdframed}
\mdfdefinestyle{figframe}{innertopmargin=5pt, linecolor=white}
\usepackage[labelfont=bf]{caption}
\usepackage{sty/commands}
\usepackage{sty/glossary}
\usepackage{hyperref}
\hypersetup{
    colorlinks,
    citecolor=black,
    filecolor=black,
    linkcolor=black,
    urlcolor=black
}
\hypersetup{colorlinks=true,
pdftitle={Weighted Delta-Tracking in Scattering Media},
pdfauthor={J.S. Rehak, L.M. Kerby, M.D. DeHart, R.N. Slaybaugh}}

\glsdisablehyper
\input{gls/gls}

\begin{document}
\begin{frontmatter}
  \title{Weighted Delta-Tracking in Scattering Media} \author[ucb]{J.~S.~Rehak\corref{cor1}}
  \ead{jsrehak@berkeley.edu} \cortext[cor1]{Corresponding author}

  \author[isu,inl]{L.~M.~Kerby} 
  \author[inl]{M.~D.~DeHart}
  \author[ucb]{R.~N.~Slaybaugh}

 \address[ucb]{University of California, Berkeley} \address[isu]{Idaho State University}
  \address[inl]{Idaho National Lab}

  \begin{abstract}
    In this work, we expand the weighted delta-tracking routine to
    include a treatment for scattering. The weighted delta-tracking
    routine adds survival biasing to normal delta-tracking, improving
    problem figure of merit. In the original formulation of
    this method, only absorption events were considered. We have
    expanded the method to include scattering and investigated the method's effectiveness with two test cases: a pressurized water reactor pin cell and a
    fast reactor pin cell. 
    We compare the figure of merit for calculating infinite flux and total
    cross-section while incrementally changing the amount of weighted
    delta-tracking used. We find that this new \gls{wdt} routine has strong potential to improve the efficiency of fast reactor calculations, and may be useful for light water reactor calculations.
    % in the abstract we need the outcome of this examination. Don't make us read the paper, tell us what you fond out. Then we'll figure out if we want to read the paper :)
  \end{abstract}


\end{frontmatter}
\section{Introduction}
\label{sec:intro}

% overall things need a little more transition and explanation. It
% feels a little hard to track what is happenening why.
%
% I tried to fix some of the flow issues, adding more information in
% transitions between sections - JSR

Monte Carlo methods have been used to model neutron transport since
first introduced at Los Alamos National Lab by Metropolis and
Ulam~\cite{metropolis1949}. Over the last 60 years, the method has
been adapted to new computer architectures and expanded to improve the
efficiency and accuracy of calculations. At the core of Monte Carlo
simulations is the statistics of simulating repeated events, so
improving the accuracy of results either requires more events or a
more efficient use of events. In transport calculations, particle
collision events are commonly and effectively used to calculate
reaction rates and fluxes.

Modern Monte Carlo codes use many techniques to improve efficiency and
in many cases operate alongside deterministic codes. These
improvements and hybrid methods have advanced the scale and types of
problems that users can simulate. The field is now seeking to solve
full-core simulations, coupled multiphysics models, and other problems on the
edge of our computational capability. One such use of multiphysics models that
seeks to support development of advanced reactor fuels is for the restart
of the \gls{treat} at the \gls{inl}~\cite{ortensi2016}. There,
researchers are using the Serpent 2 Monte Carlo
code~\cite{leppanen2007} to generate multigroup cross-sections for the
Rattlesnake deterministic code. 
% moose is a framework; rattlesnake is the deterministic code
% possibly need a reference for rattlesnake?

To advance these complex models, more efficient methods that require
less clock time are needed.  One such improvement was the introduction
of Woodcock delta-tracking~\cite{woodcock1965}. Delta-tracking
mitigates some inefficiencies that occur when standard ray tracing is
used for geometries with optically thin regions, but introduces new
well-documented difficulties in the presence of heavy absorbing
materials~\cite{leppanen2010}\cite{morgan2015}. A new method,
\gls{wdt}, was introduced by Morgan and Kotlyar~\cite{morgan2015} to
modify the delta-tracking routine to use weight reduction in place of
rejection sampling.  

% Perhaps what might make more sense is a general description of the types
% of large problems people are wating to solve. TREAT can be an example. 
% It would be motivating to discuss that these are at the edge of our computational
% capability and that we new methods like this one to be able to really solve
% these problems accurately. It's okay if this adds another paragraph.

In this paper, we will begin by providing background discussion about
the neutron propagation techniques ray tracing, delta-tracking, and
the \gls{wdt} routine of Morgan and Kotlyar. We will then describe the
new work: a novel modification of the \gls{wdt} routine that extends
the method to include scattering. This extension is a hybrid of
\gls{wdt} and normal delta-tracking that seeks to be more efficient
than either independently.  Including scattering enables us to
evaluate the \gls{wdt} method in two common test problems: a \gls{pwr}
pin cell and a fast reactor pin cell. We will then examine the results
of these simulations and compare them to a base case where the new
routine is not used. Finally, we will discuss those results and make a
recommendation for the use of \gls{wdt} in these types of problems.

% I'd add another clause or sentence here saying slightly more about the extension.

\section{Background}
\label{sec:background}

At the core of every Monte Carlo neutron transport simulation is the
propagation of the particles through material regions. Ray tracing is
a straightforward method for sampling the path length, how far a
neutron will travel before interacting with the local material.  The
algorithm samples the path length $s$ of a propagating neutron at a
position $\vec{r}$ and with energy $E$ using
\begin{align}
\label{eq:path_length}
  s(\xi, \vec{r}, E) = -\frac{1}{\Sigma_t(\vec{r},E)}\ln(\xi)\:,
\end{align}
where $\xi$ is a uniformly distributed random variable
$\xi \in [0, 1)$, and $\Sigma_t(\vec{r}, E)$ is the total macroscopic
cross-section.

The macroscopic cross-section depends on the current material, and is
therefore a discontinuous function that varies with position and the
material geometry of the problem~\cite{leppanen2013}. A path length
sampled using Eq.~\eqref{eq:path_length} becomes statistically invalid
if the neutron crosses into a material region
with a different macroscopic cross-section. Therefore, each time a new
path length is sampled, we must calculate the distance to the nearest
boundary along the propagation path. Then, if the neutron will cross
into a new material region, the distance to interaction past the
boundary must be recalculated.

These two processes cause boundaries to contribute to the
computational work of the ray tracing routine. This overhead can cause
significant inefficiencies in complex geometries or optically thin
regions. Various methods have been developed to overcome these regions
of high cost, one of which is Woodcock Delta Tracking.
% I'd end by pointing out situations where this can be costly.
% Done, do you think I need a citation? -jsr
% no -rns

\subsection{Woodcock delta-tracking}
\label{sec:delta-tracking}

Woodcock introduced the delta-tracking method for neutron propagation
in 1965~\cite{woodcock1965}; in modern Monte Carlo codes it is often
offered as an alternative or compliment to the standard ray tracing
procedure. The method was designed to address some of the limitations
of ray tracing but introduces its own new limitations.  Lux and
Koblinger~\cite{lux1991} provide an overview of the method, and its
performance in various codes has been examined by
Lepp\"{a}nen~\cite{leppanen2013} and others. A short description of the
delta-tracking method is given here to motivate development of the
\gls{wdt} routine.

The Woodcock delta-tracking method chooses a single macroscopic
cross-section to calculate path lengths for the entire region of
interest, which can include multiple materials. This is chosen to be
the maximum of all material total cross-sections, the majorant cross
section:
\begin{equation}
  \label{eq:majorant}
  \Sigma_\mathrm{maj} \equiv \max_{\vec{r} \in \set{D}}\{\Sigma_t(\vec{r})\}\:,
\end{equation}
where $\set{D}$ is the region of interest. We relate this value to the
actual macroscopic cross-section by introducing an arbitrary $\delta$
cross-section that varies with position,
\begin{equation}
  \label{eq:majorant2}
  \Sigma_\mathrm{maj} = \Sigma_\delta(\vec{r}) +
  \Sigma_t(\vec{r}), \quad\forall \vec{r} \in \set{D}\:.
\end{equation}
To maintain the physics of the problem, a delta collision must
preserve the energy and direction and existence of the neutron; it
must be a ``virtual'' collision. As a result, these can occur any
number of times along a neutron's path without consequence.

As the majorant cross-section is not a function of position, path lengths
sampled using Eq.~\eqref{eq:path_length} are valid throughout the
region of interest. To replicate sampling the real distribution that
\textit{is} a function of position, we use a rejection-sampling
algorithm, thoroughly described by Lux and
Koblinger~\cite{lux1991}. Following each sampled path length, the
collision is real (non-virtual) with a probability related to the
cross section at the point of collision,
\begin{align}
  \label{eq:preal}
  P_{\text{real}}(\vec{r}) &= \frac{\Sigma_t(\vec{r})}{\Sigma_\mathrm{maj}}\:.
\end{align}
Otherwise, the collision is a product of the $\delta$ cross-section
and is therefore virtual. In the material with the maximum
cross-section, $\Sigma_t(\vec{r}) = \Sigma_{\text{maj}}$, all
collisions are real. The algorithm for delta-tracking is shown in
Alg.~\ref{alg:dt}.
\begin{figure}[hbtp]
  \centering
  \begin{algorithm}[H]
\caption{Delta-tracking}\label{alg:dt}
\begin{algorithmic}[1]
  \State \textbf{Sample} path length, $s$

  \State \textbf{Propagate} neutron

  \State \textbf{Look up} material at neutron position to get $\Sigma_t(\vec{r})$
  \State $P_\mathrm{real} \gets
  \frac{\Sigma_t(\vec{r})}{\Sigma_\mathrm{maj}}$
  \State \textbf{Sample} random number $\xi \in [0,1)$
  \If{$\xi < P_\mathrm{real}$} 
  \State \textbf{Execute} real collision
    \Else 
    \State \textbf{Execute} virtual collision
  \EndIf
\end{algorithmic}
\end{algorithm}
\label{fig:dt}
\end{figure}

% There's an extra line break here. The journal should fix that. 
The delta tracking routine provides the ability to sample a path
length that is valid throughout the entire region of interest, unlike
ray tracing. Some of the inefficiency of ray tracing in optically thin or complex
geometries is mitigated. However, lookup is still required for the cross section at the
point of collision, and an extra random number must be sampled. 

This routine introduces its own inefficiencies. These can occur in regions where
$\Sigma_t(\vec{r}) \ll \Sigma_{\text{maj}}$ and the probability of a
real collision is low, as in streaming regions near strong
absorbers. 
% moderators usually have high scattering
In these regions, many virtual collisions occur, and the
computational expense out-weighs the little statistical information
provided. We will describe two approaches to counteract this: the
first is a combination of ray tracing and delta-tracking, and the
second is a method called \acrlong{wdt}.

\subsection{Delta-tracking with ray tracing}
\label{sec:serpent2}

Since delta-tracking can be computationally inefficient in regions
where the probability of a real collision is low, we can attempt to
mitigate this by switching back to a ray tracing routine in those
regions. An example of this approach is presented here, as it is
implemented in the Serpent 2 Monte Carlo code.

Serpent was developed at \gls{vtt} by Jaakko
Lepp\"{a}nen~\cite{leppanen2007}, and a second iteration of the code,
Serpent 2, is under active development. Both versions of Serpent use a
combination of ray-tracing and Woodcock delta-tracking for sampling
path length. Serpent selects between ray tracing and delta-tracking by
examining the ratio of total cross-section to majorant
cross-section~\cite{leppanen2010}, giving the value of
$P_{\mathrm{real}}$. In regions where many virtual collisions would
occur, the code preferentially switches to ray tracing. This is to
avoid the computational inefficiency identified in the previous
section.  This selection is determined by a constant $c$ and the
following inequality, which is identical to our formulation of
$P_{\mathrm{real}}$:
\begin{equation}
  \label{eq:s2deltasurface}
  \frac{\Sigma_t(\vec{r})}{\Sigma_\mathrm{maj}} = P_{\text{real}} > 1 - c\:.
\end{equation}
If this inequality is satisfied, delta-tracking is used, otherwise ray
tracing is used. By default, the value of $c$ is 0.9, empirically
determined to produce the largest improvement in run
time~\cite{leppanen2010}. Prior to sampling path length, the code
tests this ratio for the current neutron position and determines which
path length sampling algorithm should be used.

\subsection{\Acrlong{wdt}}
\label{sec:wdttheory}

Morgan and Kotlyar~\cite{morgan2015} introduced a different approach
to improve the inefficiencies of Woodcock delta-tracking in the
presence of large absorbers. The method, \gls{wdt}, replaces the
rejection sampling of delta-tracking with a statistical weight
reduction.

Statistical weight is an often-used mathematical concept that
describes the percentage of a neutron still available to contribute to
a response. For example, the collision estimator gives an estimate for
scalar flux $\phi$ based on the neutron's weight prior to collision
$w_i$ and the total
cross-section, summed over all collisions $C$,
\begin{align*}
  \phi = \frac{1}{W}\sum_{C}\frac{w_i}{\Sigma_t(\vec{r},E)}\:,
\end{align*}
where $W$ is the weight of neutrons at generation, which is often unity.
% I'd put in an equation for a tally here to provide concrete context. 
A neutron with zero weight is removed from the simulation, because it
can no longer contribute to responses.

Neutron weight can
be augmented throughout its lifetime through interactions, games
played at boundary crossings, etc., so long as a fair game is
maintained. Weight change methods such as survival biasing improve the
efficiency of simulations by keeping particles alive longer, allowing them to
contribute more information to the solution with the goal of improving statistics. 
% somehow the idea of hope and science feels weird to me.
Details of different methods for variance reduction are outside the
scope of this work. The main points are that weight corresponds to
contribution to the solution of interest, weight can be changed based
on a particle's history, and a fair game must be maintained.

% Hopefully what I just changed this to is what you were going for...

The \gls{wdt} method samples the particle path length in the same
fashion as Woodcock delta-tracking, using the majorant
cross-section. Unlike the delta-tracking routine, \gls{wdt} accepts
all collisions as real. To keep a fair game, the neutron's weight must
be changed to account for the virtual collisions that no longer occur
explicitly. Each collision can result in multiple outcomes, each of
which is mutually exclusive. To calculate the change in weight, we use
the expected value of the interaction.
% we haven't described implicit capture. Is that supposed to be above here?
The outcomes are real and virtual collisions, which result in final
weights $w_{f, \mathrm{virt}}$ and $w_{f, \mathrm{real}}$,
respectively. The final neutron weight is therefore given by
% please define w_{f,\mathrm{real}} and w_{f,\mathrm{virt}} before using
\begin{equation}
  \label{eq:wdtexpected}
  E[w_f] = w_{f,\mathrm{real}}P_{\mathrm{real}} +
  w_{f,\mathrm{virt}}\left(1 - P_{\mathrm{real}}\right)\:.
\end{equation}
An absorption event removes the particle from the simulation, so the
resulting final weight of a real collision is zero. A virtual
collision is a scattering event, and therefore leaves the weight
unchanged. Inserting the appropriate values into
Eq.~\eqref{eq:wdtexpected} gives the expected value of the final
weight for an absorption event for a neutron with initial weight $w_i$:
% also define w_i
\begin{align*}
  \label{eq:mkexpected}
  E[w_f, \vec{r}] &= w_{f,\mathrm{real}}P_{\mathrm{real}}(\vec{r}) +
           w_{f,\mathrm{virt}}(1 - P_{\mathrm{real}}(\vec{r})) \\
  &= w_i\left(1 - P_{\mathrm{real}}(\vec{r})\right)\:.
\end{align*}
The particle then continues propagating as if it underwent a virtual
collision. The removed statistical weight is then added to the
appropriate responses, such as collision tallies. %This is shown
%schematically in Fig.~\ref{fig:wdt_schematic}.
% also please define P_{virt}
% I tried to keep P_{virt} out of it, I'm removing it from the next
% section. If you think it should be in there,  I can add it, it just
% seemed superfluous. I feel like 1-P_{real} gets the point across -jsr
% ok -rns

% \begin{figure*}[hbtp]
%   \centering
%   \includegraphics[scale=0.4]{img/wdt}
%   \caption{WDT routine for an absorption event of an incoming neutron
%     with incident weight, $w_i$, energy $E_i$, and velocity
%     $\vec{v}_i$}
% \label{fig:wdt_schematic}
% \end{figure*}

The key change in \gls{wdt} is that unlike normal delta-tracking,
\textit{every} sampled path length results in a real collision and
subsequent contribution to tallies. As discussed earlier, a major
inefficiency in delta-tracking is wasted computational resources in
regions with many virtual collisions that do not contribute to
statistical quality. The \gls{wdt} method therefore partially
mitigates this issue by collecting contributions to tallies with every
collision.

% Statement about why we would want that and what the expected
% behavior would be

This algorithm was implemented by Kotlyar and Morgan and tested using
a 1D problem in an absorbing medium. They found an improvement in the
computational efficiency of the simulation.

Legrady, Molnar, Klausz, and Major~\cite{Legrady2017} performed an
analysis of delta-tracking methods, including \gls{wdt}. The aim of
their study was to quantify the effects of modifying the single
cross-section used by the Woodcock delta-tracking method for sampling
path lengths. This sampling cross-section is chosen to be the majorant
in standard Woodcock delta-tracking and \gls{wdt}. They find that the
optimal sampling cross-section occurs below the majorant in all cases
and outperformed the \gls{wdt} method by a factor of 70 in their test
cases. For this study, we will not consider their optimization, and
will leave the sampling cross-sections of the standard delta-tracking
and \gls{wdt} processes unmodified.

\section{Method}
\label{sec:method}

The \gls{wdt} routine introduced a method to improve the
inefficiencies of Woodcock delta-tracking in the presence of large
absorbers. The method resulted in an improvement in simulation
efficiency in a purely absorbing medium, but did not address how the
routine would work with scattering. Therefore, we will need to extend
or modify the \gls{wdt} method to include scattering such that it is
applicable to more realistic problems.

We will first attempt to extend \gls{wdt} in a na\"{i}ve fashion, by
applying the same process for absorption to scattering. We will show
that this will result in a process that can become intractable. This
will motivate development of a hybrid method that uses both standard
delta-tracking and \gls{wdt} to accommodate scattering events.

% I think we need some clearer explanation of where we're going. Let
% us know that we're going to attempt a straightforward extension for
% the purposes of illustrating why we need the approach actually
% taken. I think that will help frame it, and then the next few
% paragraphs need some explanation.

\subsection{Na\"{i}ve Extension of \Acrlong{wdt}}
\label{sec:wdt_scattering}

As previously discussed, in a scattering event the statistical weight
of the incident particle does not change. This holds true for a real
scattering event as well as a virtual event; from the perspective of
neutron weight, the two events are identical. Therefore, application of
the \gls{wdt} weight reduction will result in \textit{no} overall
weight reduction:
\begin{align*}
  E[w_f] &= w_{f,\mathrm{real}}P_{\mathrm{real}} +
           w_{f,\mathrm{virt}}(1 -P_\mathrm{real}) \\
  &= w_i(P_\mathrm{real} + 1 -P_\mathrm{real}) \\
  &= w_i\:.
\end{align*}
Although the overall statistical weight remains unchanged, portions of
it must be added to the appropriate response. The \gls{wdt} weight
reduction splits the statistical weight into two portions: the real
portion ``experiences'' the scattering event, and the virtual portion
undergoes a virtual collision. The virtual portion of the weight is
retained by the original neutron, which continues with the same
direction and energy. We then create a new neutron to receive the real
portion and execute a real scattering
event.
%Fig.~\ref{fig:wdt_schematic_scattering} shows a schematic of
%this process.

% I am super confused by the previous paragraph.
% We started with scatteringbut then described what happens with absorption
% then the next paragraph describes  scattering?
% , as before, the neutron retains

% Reworded, I removed discussion of absorption - jsr
% I think this is clear now -rns

% \begin{figure*}[hbtp]
%   \centering
%   \includegraphics[scale=0.4]{img/wdt_scattering}
%   \caption{WDT routine for a scattering event of an incoming neutron
%     with incident weight, $w_i$, energy $E_i$, and velocity
%     $\vec{v}_i$}
% \label{fig:wdt_schematic_scattering}
% \end{figure*}

Therefore, straightforward extension of this methodology to scattering
requires doubling of the neutron at every scattering event. In
problems with any scattering, this will likely result in a rapid and
intractable multiplication of neutrons that will overload memory buffers. 
In an experimental implementation of this method, that is exactly what we found. 
Thus, we concluded the na\"{i}ve extension was not a useful way to 
include scattering in \gls{wdt}.

\subsection{Hybrid Extension of \Acrlong{wdt}}
\label{sec:scattering}
% or some title change of the previous one and this one that are set up by a section introductory paragraph. 

\begin{figure}[hbtp]           
  \includegraphics[scale=0.4]{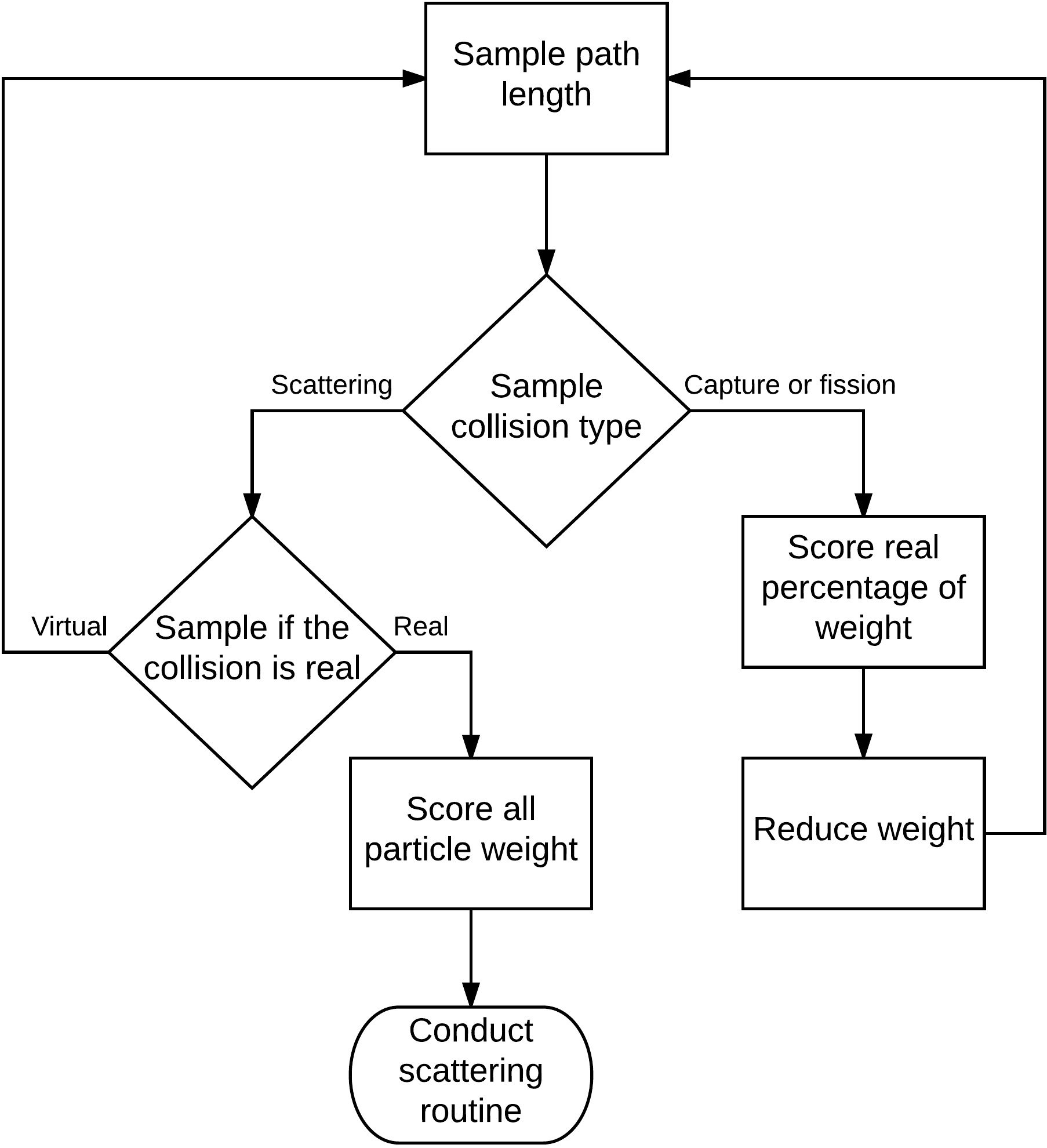}

    \caption{\Acrlong{wdt} with scattering rejection
                  sampling.\label{fig:wdt}}
\end{figure}

To address the issue of intractable neutron multiplication that comes
with the na\"{i}ve extension of \gls{wdt}, we created a hybrid version
that includes a fallback to the standard delta-tracking method for any
scattering events. Following each sampled path length, the type of
collision (that will occur if real) must be sampled; we use \gls{wdt}
if the collision is an absorption event, and delta-tracking
otherwise. The algorithm is shown in Alg.~\ref{alg:scattalg} and a
flow chart of the routine is shown in Fig.~\ref{fig:wdt}.

\begin{algorithm}[hbtp]
\caption{\Acrlong{wdt} with scattering}\label{alg:scattalg}
\begin{algorithmic}[1]
  \State \textbf{Sample} path length
  \State \textbf{Sample} collision type
  \If{collision type == (capture or fission)}
    \State \textbf{Score} capture or fission $\gets
    w_iP_\mathrm{real}$
    \State \textbf{Score} collision  $\gets w_iP_\mathrm{real}$
    \State $w_f \gets w_i(1-P_\mathrm{real})$
    \State \textbf{Execute} virtual collision
  \Else
    \State \textbf{Sample} random number $\xi \in [0,1)$
    \If{$\xi < P_\mathrm{real}$} \Comment{Collision is real}
    \State \textbf{Score} scattering $\gets w_i$
    \State \textbf{Score} collision $\gets w_i$
      \State \textbf{Execute} scattering collision
    \Else \Comment{Collision is virtual}
    \State \textbf{Execute} virtual collision
    \EndIf
  \EndIf
\end{algorithmic}
\end{algorithm}

In highly scattering regions, the routine is expected to be slightly
less efficient than standard delta-tracking because every collision
requires an extra random number sample and comparison. By contrast, in
highly absorbing regions, we aim to benefit from the improved
efficiency of the \gls{wdt} routine compared to delta-tracking.

The efficiency will also depend on the value of $P_{\text{real}}$. At
high values of $P_{\mathrm{real}}$, a majority of the weight of the
incoming particle is scored. This leaves the particle that undergoes a
virtual collision with a very low weight, relying on a rouletting
routine~\cite{lux1991} to prevent computational inefficiency. 
% need a citation since we didn't describe rouletting.
% Went with the Lux textbook on MC, because they give a thorough
% description -jsr
In regions of low $P_{\text{real}}$, each collision will only
contribute a small amount to the appropriate tallies. We expect this
to be more efficient than standard delta-tracking, which generates
many virtual collisions in the same region that do not contribute to
tallies of interest.

This implies
that there is a region between low and high values of
$P_\mathrm{real}$ where \gls{wdt} may provide benefit.

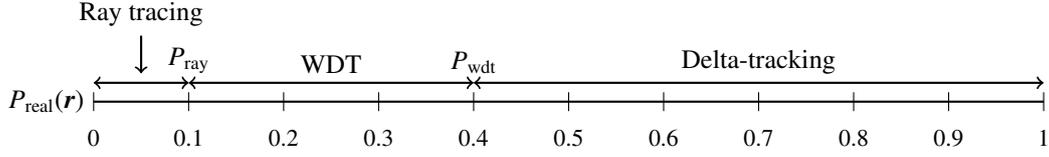
\begin{figure*}[h]
  \centering
  \begin{tikzpicture}[scale=1.25]
    \draw[thick] (0,0) -- (10.0,0);
    \foreach \x in {0,1,...,10}
    {
      \draw (\x, 0.1) -- (\x, -0.1);
      \pgfmathsetmacro\result{\x * 0.1}
      \node [below] at (\x, -0.2) {\small $\pgfmathprintnumber{\result}$};
    }
    \node [left] at (0,0) {$P_{\mathrm{real}}(\vec{r})$};
    \draw[ thick, <->] (0,0.2) -- (1,0.2);
    \draw[ thick, <->] (1,0.2) -- (4,0.2);
    \draw[ thick, <->] (4,0.2) -- (10,0.2);
    % \node [above] at (0.5, 0.2) {Ray tracing};
    \node [above] at (0.5, 0.7) {Ray tracing};
    \draw[ thick, ->] (0.5, 0.7) -- (0.5, 0.3);
    \node [above] at (2.5, 0.2) {WDT};
    \node [above] at (7, 0.2) {Delta-tracking};
    \node [above] at (1, 0.2) {$P_\mathrm{ray}$};
    \node [above] at (4, 0.2) {$P_\mathrm{wdt}$};
  \end{tikzpicture}
  \caption[Implemented selection scheme for ray-tracing, weighted and normal
    delta-tracking.]{Implemented selection scheme for propagation
      mode: ray-tracing, weighted, and normal
    delta-tracking. Shown using the values of $P_{\mathrm{ray}}=0.1$ and $P_{\mathrm{wdt}}=0.4$.}
  \label{fig:ray_wdt}
\end{figure*}

We have previously discussed that switching to a ray tracing routine
in regions of low $P_\mathrm{real}$ is effective in countering the
inefficiency of delta-tracking. We therefore combine ray tracing,
\gls{wdt}, and standard delta-tracking together in a scheme defined by
two parameters. On the lower end,
the value of $P_{\mathrm{ray}}$ defines the value of $P_\mathrm{real}$ below which ray
tracing is used. On the upper end, $P_{\mathrm{wdt}}$ defines the
cutoff below which \gls{wdt} will be used instead of normal
delta-tracking. This scheme is summarized as,
  \begin{align*}
  &\text{Propagation mode }(\vec{r}, E) =\\&\left\{
  \begin{array}{lrrcl}
      \text{Ray tracing}, & \text{if} & & P_{\mathrm{real}}(\vec{r}, E) & <
                                                           P_{\mathrm{ray}} \\
    \text{\gls{wdt}}, & \text{if} & P_{\mathrm{ray}} \leq &
                                                P_\mathrm{real}(\vec{r}, E)
                                                         & <
                                                           P_{\mathrm{wdt}}\\
                                                           \text{Delta-tracking}, & \text{if}
     &   P_{\mathrm{wdt}}\leq & P_\mathrm{real}(\vec{r}, E) &
  \end{array}\right.\:,
  \end{align*}
  where $P_\mathrm{real} \in [0, 1)$ and is shown graphically in
  Fig.~\ref{fig:ray_wdt}.

Implementation in Serpent was straightforward; $P_{\mathrm{ray}}$ is
already implemented in the user parameter $c$, where
$P_{\mathrm{ray}} = (1-c)$. As discussed in Sec.~\ref{sec:serpent2}, a
value of $c = 0.9$ or $P_{\mathrm{ray}} = 0.1$ resulted in the highest
computational efficiency. Examination of the \gls{wdt} cutoff value,
$P_{\mathrm{wdt}}$, is the aim of the remainder of this study. We
implemented our hybrid \gls{wdt} in Serpent 2, with the ability to
adjust the value of $P_{\mathrm{wdt}}$. The amount of ray tracing
never changes; we are instead adjusting the amount of delta-tracking
that is replaced by the \gls{wdt} routine. By adjusting the value of
$P_\mathrm{wdt}$ from
$P_\mathrm{ray}$ (no \gls{wdt}) to 1.0 (full \gls{wdt}), we investigate the impact
of the routine on computational efficiency.
% what about adding t_wdt?

% you way want to add tray and twdt as markers
% also, this is too wide for the page :)

\section{Results \& Discussion}
\label{sec:results}

We expect that the \gls{wdt} method will improve the statistics of
Serpent calculations. With normal delta-tracking, virtual collisions
provide no statistical benefit, and are therefore an inefficient use
of computational resources. In contrast, all collisions that result in
absorption events contribute to statistics when using
\gls{wdt}. Therefore, we expect an improvement in the results of a
Serpent 2 simulation in regions with absorption. We chose two test
cases: a fast reactor pin cell and a PWR pin cell. In this section,
we will describe those two cases, how we determined quality in the
simulation, and the test case results.

\subsection{Test Cases}\label{sec:test_cases}

Two test cases were selected to test the performance of the \gls{wdt}
method. These are a \gls{pwr} pin cell and a fast reactor pin cell. We
chose these two test cases to evaluate the method in a domain where
absorption is dominant in the fast group (fast pin cell) and one in
which absorption is dominant in the thermal group (\gls{pwr} pin cell).
% I don't understand what this sentence means. Most of the scattering is in the thermal group, right?
% how much scattering?
% Reworked this section, I think the main point is that each has a
% different regime of absorption. -jsr
% This is easier for me to understand now -rns
All test cases were run on a small cluster at the University of
California, Berkeley, and the specifications of the cluster are shown
in Tab.~\ref{tab:abacus}.
% Either here or at the begining of each test case description, I'd
% add a few sentences, up to a paragraph, about why each is a relevant
% test problem and what you expect to learn from it. That primes us
% for what to look for in the data and why we care about these tests
% in particular. This will also help us understand the boundaries of
% how tested the method is.
\begin{table}[hbtp]
  \centering
  \caption{Small cluster specifications}
  \begin{tabularx}{\columnwidth}{lX}
    \toprule
    \textrm{Parameter} & \textrm{Specification} \\ \midrule
    Processor & 2 $\times$ TenCore Intel Xeon Processor E5--2687W v3
                3.10 GHz 25MB Cache \\
    RAM & 16 $\times$ 16GB PC4--17000 2133MHz DDR4 \\
    Hard drive & 2 $\times$ 800 GB Intel SATA 6.0 GB/s Solid State Drive
          \\ \bottomrule
  \end{tabularx}\label{tab:abacus}
\end{table}

\subsubsection{\Acrlong{pwr} pin cell}\label{sec:pwr}
\begin{figure}[hbtp]
  \centering
  \begin{tabular}{cc}
  \includegraphics[scale=0.25]{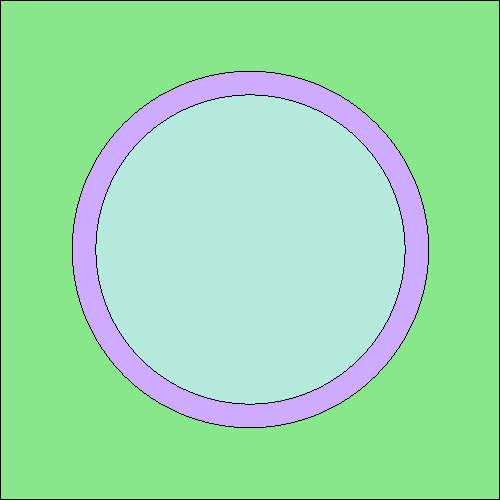} \\
\begin{tabular}{@{}lr@{}}
\toprule
\textbf{Parameter}     & \textbf{Value (cm)} \\ \midrule
Fuel radius     & 0.412          \\
Cladding radius & 0.475          \\
Pitch              & 1.33          \\ \bottomrule
\end{tabular}
  \end{tabular}
  \caption[Pressurized water reactor geometry and physical
  parameters.]{Pressurized water reactor geometry and physical
    parameters. The fuel is shown in light blue, the cladding in
    purple and the coolant is green.}
  \label{fig:pwr}
\end{figure}

The \gls{pwr} pin cell was chosen from the Serpent 2 validation input
files provided on the \gls{vtt} Serpent
webpage~\cite{serpent_web}. The geometry and physical parameters are
shown in Fig.~\ref{fig:pwr}. The fuel is a 2.68 w/o enriched UO$_2$
mixture, with Zircaloy cladding, light water moderation, and
reflective boundary conditions. The simulation output statistics
binned neutrons into two energy groups with the group boundary at
0.625 eV.

\subsubsection{Fast reactor pin cell}
\label{sec:fast_pin_cell}

We adapted the fast reactor pin cell from an example provided in
the Serpent validation files~\cite{serpent_web}. This is a lead cooled
pin cell with \gls{mox} fuel containing uranium, plutonium, and a small
amount of americium. The relative isotope amounts are shown in Tab.~\ref{tab:fast_fuel}.
The cladding is stainless steel, the lattice is hexagonal, and we used 
reflective boundary conditions. As with the \gls{pwr} pin cell, we ran
the simulation with two energy groups with the group boundary at 0.625
eV.

\begin{figure}[hbtp]
  \centering
  \begin{tabular}{cc}
  \includegraphics[scale=0.25]{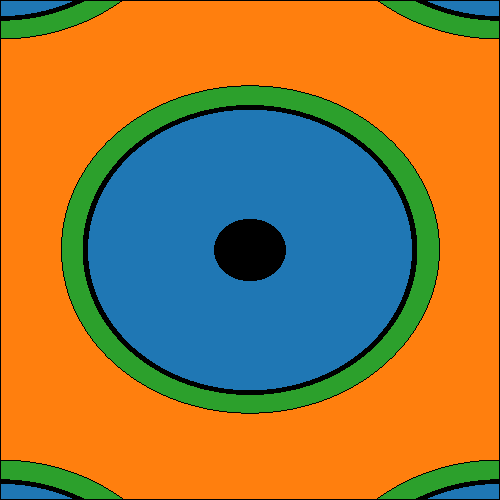} \\
\begin{tabular}{@{}lr@{}}
\toprule
\textbf{Parameter}     & \textbf{Value (cm)} \\ \midrule
Void radius     & 0.1          \\
Pellet radius & 0.45          \\
Inner cladding radius & 0.465          \\
Outer cladding radius & 0.525          \\
Pitch              & 1.789          \\ \bottomrule
\end{tabular}
  \end{tabular}
  \caption[Fast pin cell geometry and physical parameters.]{Fast pin
    cell geometry and physical parameters. The pellet is shown in
    blue, the cladding green, and the coolant orange. Black regions
    are voids.}
  \label{fig:fast}
\end{figure}

\begin{table}[hbtp]
  \caption{Atomic density of the fast reactor pin cell \acrshort{mox}
    fuel, values are relative to the atomic density of \ele{U}{238}}
  \centering
  \begin{tabular}{cr}\toprule
    Isotope & Relative Atomic Density \\ \midrule
    \ele{U}{238} & 1.00 \\
    \ele{Pu}{239} & 0.16 \\
    \ele{Pu}{240} & 7.40 $\times 10^{-2}$ \\
    \ele{Pu}{242} & 2.09 $\times 10^{-2}$ \\
    \ele{Pu}{238} & 6.45 $\times 10^{-3}$ \\
    \ele{U}{235} & 4.11 $\times 10^{-3}$ \\
    \ele{Am}{241} & 3.57 $\times 10^{-3}$ \\
    \ele{U}{236} & 1.00 $\times 10^{-4}$ \\
    \ele{U}{234} & 3.01 $\times 10^{-5}$ \\ \bottomrule
  \end{tabular}
  \label{tab:fast_fuel}
\end{table}
% is this normalized to 100% U-238? It seems weird to me that it adds
% to more than 1. Maybe I'm not quite understanding...
% Yes, I'll make that more explicit.

\subsection{Cycles per CPU time}\label{sec:cyclecpu}

All calculations done in this study were criticality source
calculations. For these types of calculations in Serpent, source
neutrons are run in cycles. Each cycle contains the same number of
source neutrons, and the source distribution is determined by the
previous cycle~\cite{leppanen2015}. More efficient simulations will be
capable of running more cycles, $C$, in a given CPU time frame, $t$. So, to
measure one standard of efficiency, we looked at the cycles per CPU
time, $C/t$, for each of the two test cases. 

% I think it would be good to have a paragraph here talking about real
% collisions per track and the relationship to cycles per CPU. You
% have it below, but I think it logically goes here more easily.
%
% Moved -jsr
% thanks. However, this section no longer made sense. I think i fixed it... -rns

Another way to consider method effectiveness is using the average number of real collisions per track. Unlike virtual
collisions, real collisions contribute to the statistics of the
problem and therefore require tallying and other processes. This makes real collisions more computationally costly than virtual collisions. Based on the \gls{wdt} and standard delta tracking algorithms, we
expect to see an increase in the number of overall real collisions
with a subsequent decrease in cycles per CPU time. This apparent
reduction in efficiency will be offset by the improved statistics
provided by the real collisions. We therefore expect cycles per CPU time and average real collisions per track to have inverted behaviors. 
% state the behavior you expect so we're primed for it
% added -jsr

\subsubsection{Fast Reactor Pin Cell}\label{sec:cyclescpu_fast}

\begin{figure*}[hbtp]
    \begin{subfigure}[t]{0.5\textwidth}
      \centering
      \includegraphics[width=\textwidth]{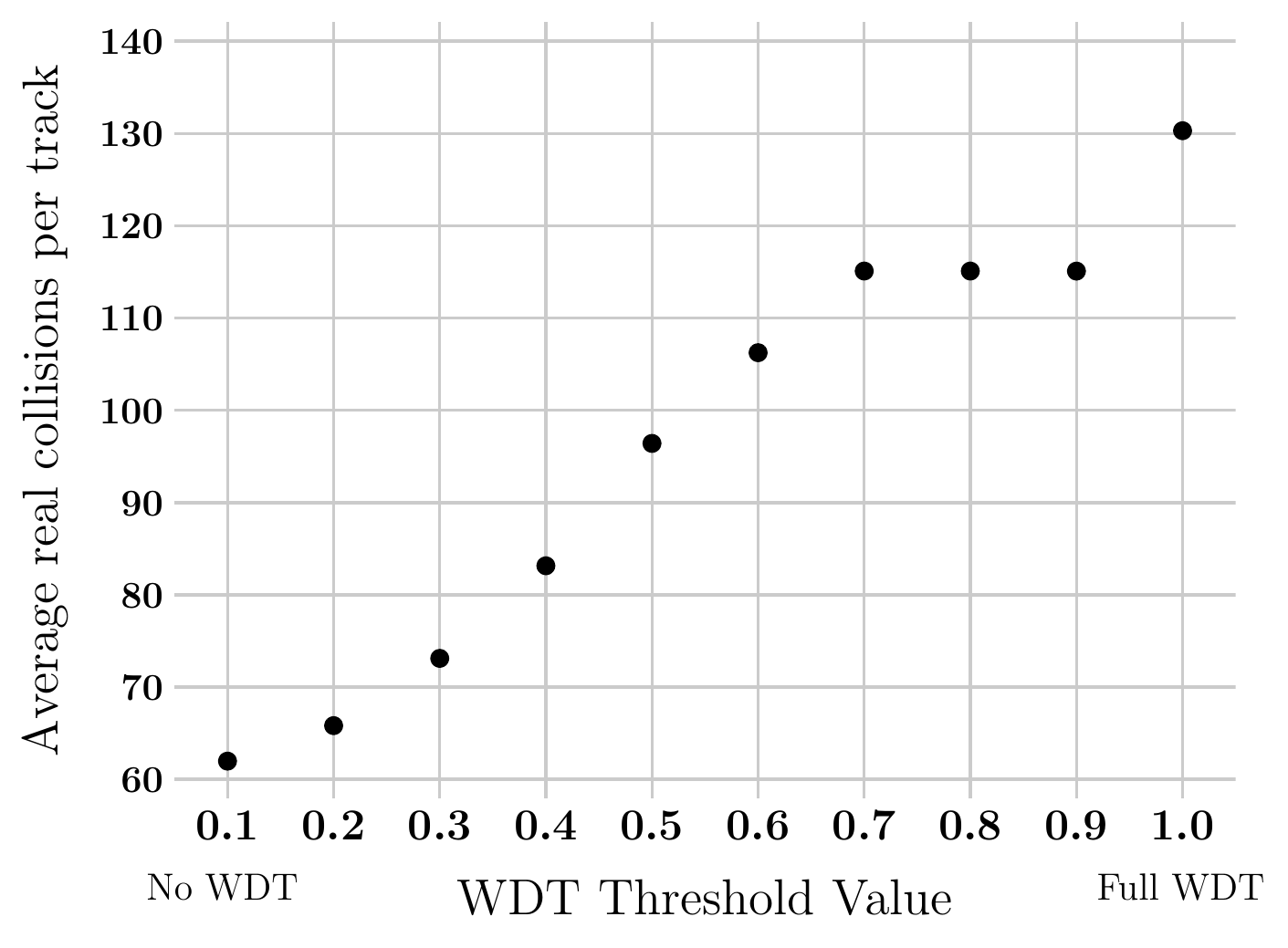}
      \caption{}\label{fig:fast_real_collisions}
    \end{subfigure}
    ~
  \begin{subfigure}[t]{0.5\textwidth}
      \centering
      \includegraphics[width=\textwidth]{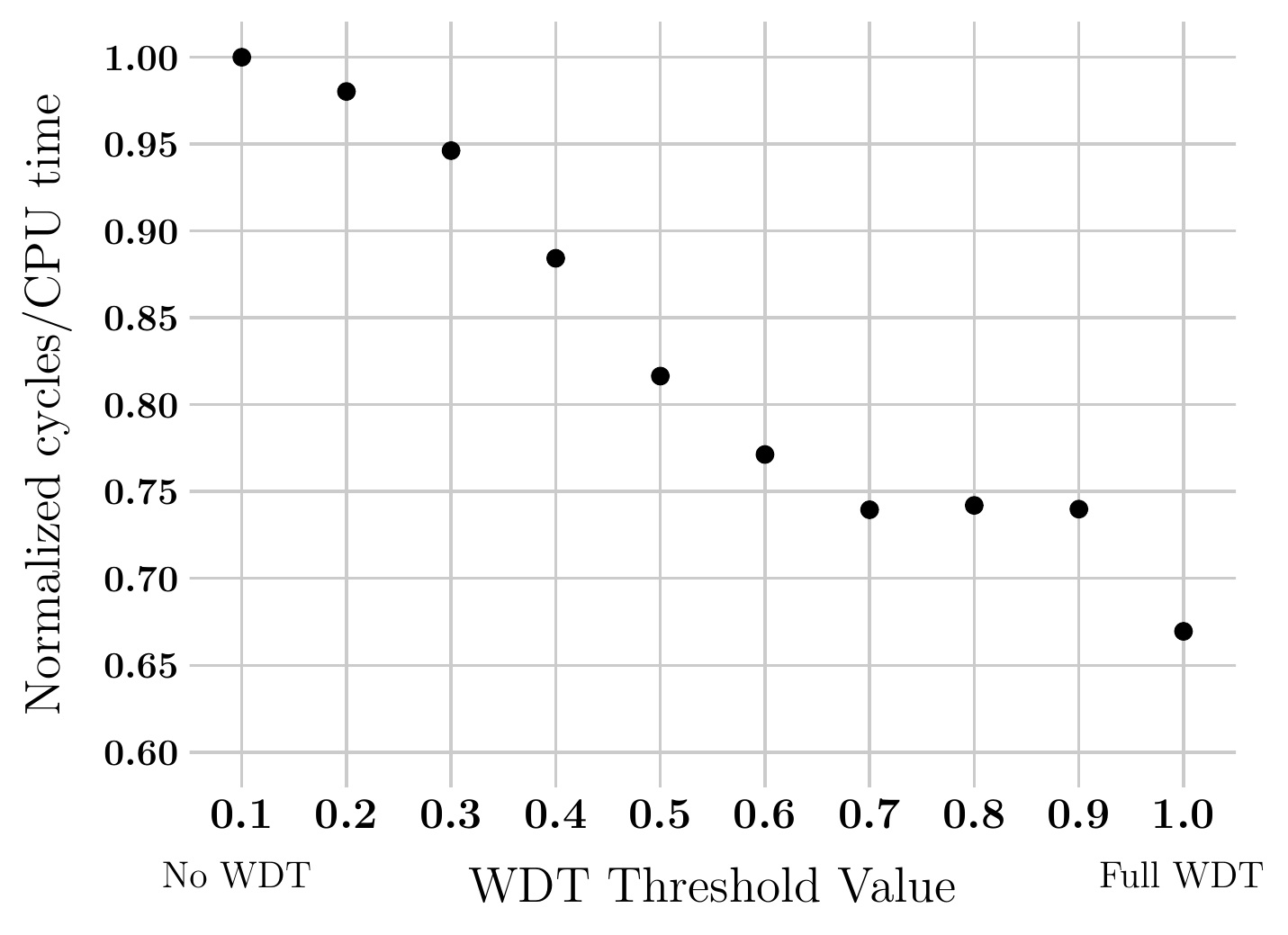}
      \caption{}\label{fig:fast_cyc_cpu}
    \end{subfigure}
    \caption{For the fast reactor pin cell, (a) cycles per CPU time
      (normalized to the base case with no \gls{wdt}) and (b) average
      real collisions per track, with increasing threshold to wdt,
      $P_{\mathrm{wdt}}$}\label{fig:cycles_cpu_real_col_fast}
\end{figure*}

The fast reactor pin cell is dominated by absorption; fast neutrons
drive fission and little scattering occurs. 
% Then here you can dive in the the results now that we know what
% we're going to be looking at and how they're related.
Our \gls{wdt} with scattering routine always considers absorption
reactions to be real collisions, so we see a rise of the average real
collisions per neutron track compared to the standard delta-tracking routine.
% standard wdt? no wdt? both? Standard Delta-tracking. -JSR
This is shown in Fig~\ref{fig:fast_cyc_cpu}. At higher levels of the
threshold value the number of real collisions levels off, indicating that the
\gls{wdt} routine may be handling a majority of the phase space where
absorption occurs. We observe a step increase in average real
collisions per track when $P_{\mathrm{wdt}}$ is unity; this is the
only value where the \gls{wdt} routine will be used in the region that
defines $\Sigma_{\mathrm{maj}}$. For this test problem the region of
highest cross-section is also a highly absorbing region, so use of the
\gls{wdt} routine in that region causes the number of real collisions
to increase significantly. We see the inverse pattern in the cycles
per CPU time, as anticipated and described in the previous
section. The data are shown in the second column of
Table~\ref{tab:cycles_cpu}.

\subsubsection{PWR Pin Cell}\label{sec:cyclescpu_pwr}

\begin{figure*}[hbtp]
  \begin{subfigure}[t]{0.5\textwidth}
      \centering
      \includegraphics[width=\textwidth]{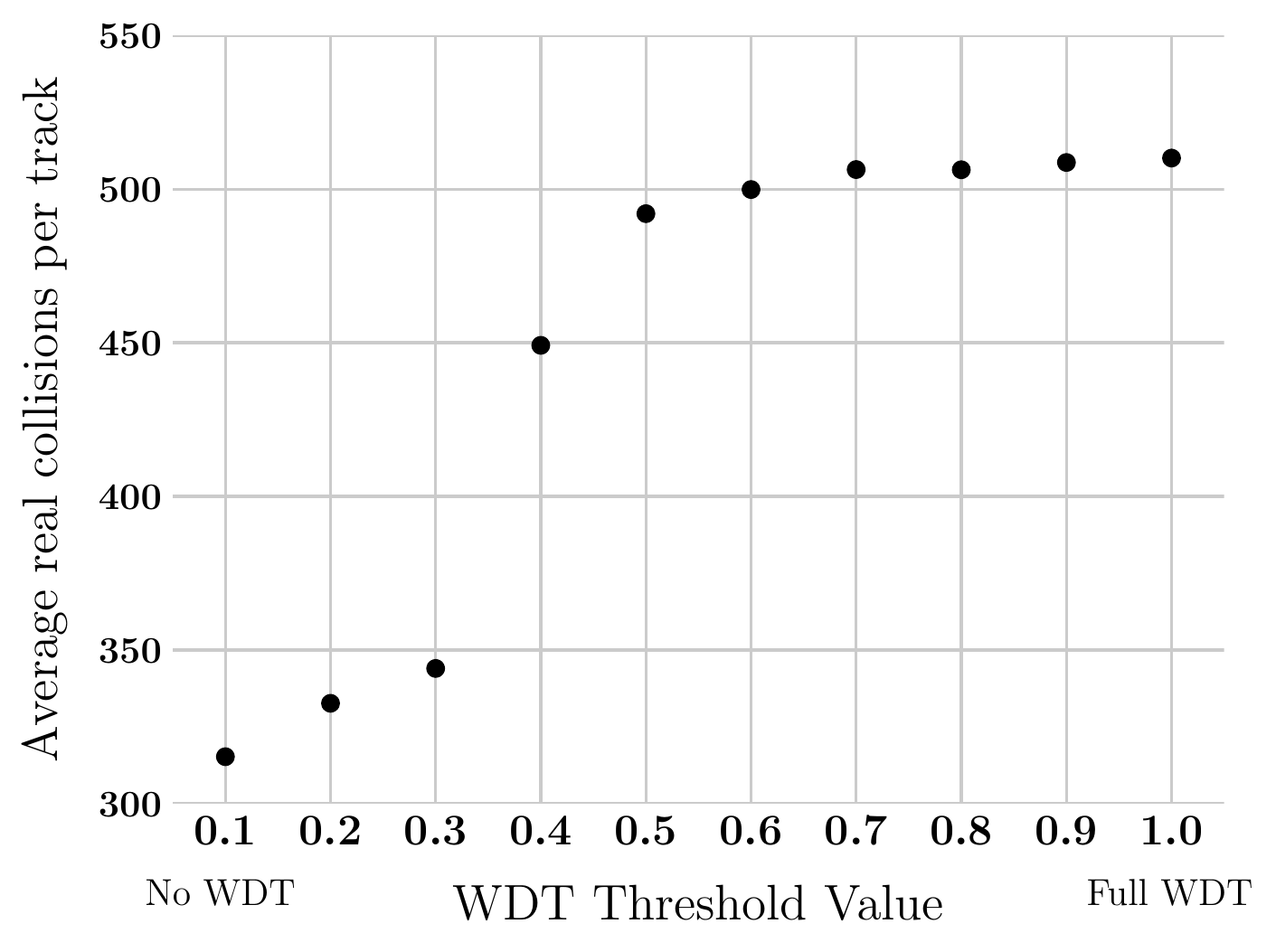}
      \caption{}\label{fig:pwr_real_collisions}
    \end{subfigure}
    ~\begin{subfigure}[t]{0.5\textwidth}
      \centering
      \includegraphics[width=\textwidth]{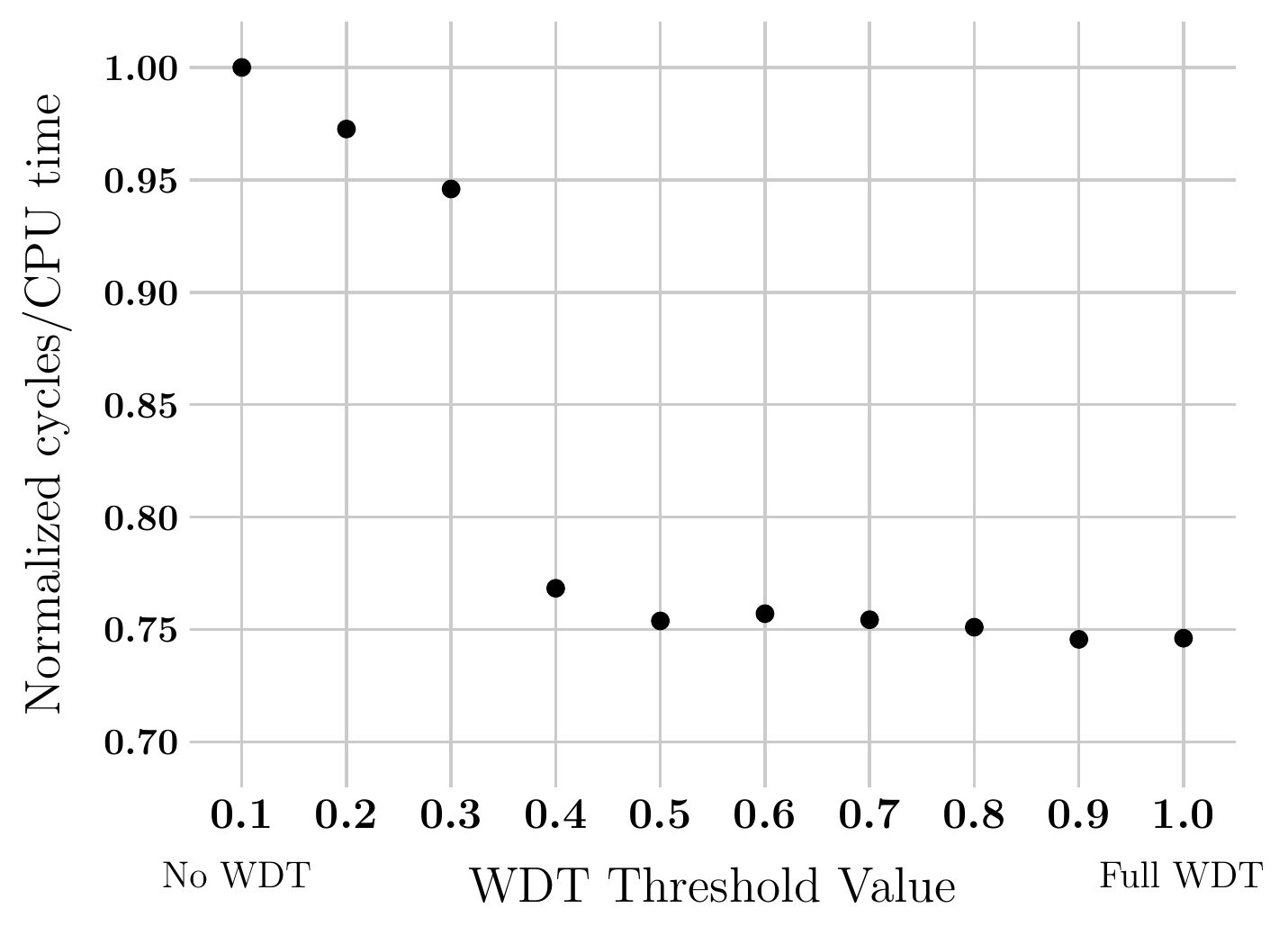}
      \caption{}\label{fig:pwr_cyc_cpu}
    \end{subfigure}
    \caption{For the PWR reactor pin cell, (a) cycles per CPU time
      (normalized to the base case with no \gls{wdt}) and
    (b) average real collisions per track, with increasing threshold
    to wdt, $P_{\mathrm{wdt}}$}\label{fig:cycles_cpu_real_col_pwr}
\end{figure*}

Unlike the fast pin cell, the PWR pin cell is dominated by
scattering. Compared to the fast pin cell, each given track has many
more real collisions, as seen in Fig.~\ref{fig:pwr_real_collisions},
and there is an exponential rise as more \gls{wdt} is introduced. As
the neutrons scatter and thermalize, the absorption cross-section for
the moderator rises exponentially leading to exponentially more
absorption events. As \gls{wdt} considers all absorption events to be
real collisions, this leads to a subsequent exponential rise in the
average real collisions per track. At higher levels of threshold,
\gls{wdt} handles the entire phase space where this benefit is
realized and the number of real collisions levels off. The
discontinuity may be resolved with finer threshold values, or may be
the result of inclusion of different materials at discrete threshold
values. We again see the expected inverse pattern in the cycles per
CPU time. The data are shown in the first column of
Table~\ref{tab:cycles_cpu}.

% I'd put real collisions on the left and cycles on the right b/c
% that's how you talk about them.
% Swapped -JSR

\begin{figure}
    \caption{Normalized cycles per CPU time for the PWR pin cell and
    fast reactor pin cell. Values are noramlized to the case with no
    WDT.}\label{tab:cycles_cpu}
  \centering
  \begin{tabular}{rrr}\toprule
    & \multicolumn{2}{c}{Cycles/CPU time} \\
    $P_\text{wdt}$ & Fast pin cell& PWR pin cell\\
    % \cmidrule(lr){2-2} \cmidrule(lr){3-3}
    \midrule
    0.1 &          1.00 &         1.00 \\
    0.2 &          0.98 &         0.97 \\
    0.3 &          0.95 &         0.95 \\
    0.4 &          0.88 &         0.77 \\
    0.5 &          0.82 &         0.75 \\
    0.6 &          0.77 &         0.76 \\
    0.7 &          0.74 &         0.75 \\
    0.8 &          0.74 &         0.75 \\
    0.9 &          0.74 &         0.75 \\
    1.0 &          0.67 &         0.75 \\    \bottomrule
  \end{tabular}
\end{figure}

\subsection{Figure of Merit}\label{sec:fom}

% I think this section needs to be reframed, again the logical flow feels a little in reverse to me.
% I'd start with the common FOM, relating time to cycles, and explaining that 
% VR is always a tradeoff between time and number of cycles. Then I think the
% rest of the dicussion I think flows a bit more logically from
% there.

% I reworked this section a bit -JSR
%

At first glance, the \gls{wdt} method appears to be less efficient
by causing more real collisions, and therefore requiring more clock
time for the simulation to run the same number of
particles. Ultimately, though, running many particles is not the goal
of the simulation. Our goal is calculating parameters of interest with
low error, so this must be taken into account. Because variances are
inversely proportional to the number of neutron cycles simulated with
this relationship,
\begin{equation}
\label{eq:variance}
  \sigma^2(\hat{x}) = \frac{C}{n}\:.
\end{equation}
where $n$ is the number of cycles and $C$ is a constant, we can
theoretically always reduce the variance at the expense of more cycles
and CPU time. The goal of efficient variance reduction is therefore to
reduce the constant $C$, such that a smaller variance can be achieved
for a given number of cycles $n$. We will describe this constant
using a \gls{fom}.

The commonly-used formulation of this \gls{fom} is described
by Lewis and Miller~\cite{lewis1993},
\begin{equation}
  \label{eq:fom}
  \mathrm{FOM} = \frac{1}{\sigma(\hat{x})^2t}\:,
\end{equation}
where $t$ is the runtime of the simulation and is proportional to the
number of cycles $t \propto n$. Plugging this and
Eq.~\eqref{eq:variance} into Eq.~\eqref{eq:fom} yields:
\begin{equation*}
  \mathrm{FOM} = \frac{1}{\sigma(\hat{x})^2t} =
  \frac{n}{C_1}\cdot\frac{1}{C_2\cdot n} = C_3\:,
\end{equation*}
where $C_1,C_2,$ and $C_3$ are constants. We therefore expect the \gls{fom}
to be a constant value independent of the number of cycles $n$.  A
higher \gls{fom} indicates lower variance per computation time, and
therefore a more efficient algorithm.

A Serpent simulation was run for both test cases using values of the
\gls{wdt} threshold ($P_\mathrm{wdt}$) from 0.1 to unity in increments
of 0.1, maintaining a constant threshold to ray tracing,
$P_{\mathrm{ray}}$, as described in Fig.~\ref{fig:ray_wdt}. We
normalized the final \gls{fom} for each value of $P_{\mathrm{wdt}}$ to
the final \gls{fom} with no \gls{wdt}:
\begin{equation*}
  \mnfom(P_{\mathrm{wdt}}) =
  \frac{\mathrm{FOM}(P_\mathrm{wdt})}{{\mathrm{FOM}}_0}\:,
\end{equation*}
where ${\mathrm{FOM}}_0 =
{\mathrm{FOM}}(0.1)$ is
the \gls{fom} with no \gls{wdt}. Increasing values of
$P_{\mathrm{wdt}}$ leads to increasing the amount of \gls{wdt} over
regular delta-tracking. The parameters for the rouletting scheme are
kept constant, with a weight cutoff of $w = 0.1$ and probability of
rouletting $P_{\mathrm{kill}} = 0.5$.

First, all simulations were run for a long enough period for the
\gls{fom} to converge. The cycles/CPU time is observed to ensure
computer loading did not change. Then, the \gls{fom} at each point in
the simulation is calculated using the total \textit{cycles}, $C$:
\begin{equation}
  \label{eq:fomc}
  \mfom_C = \frac{1}{\sigma(\hat{x})^2C}\,.
\end{equation}
Then, all connections were
isolated to the cluster so that no other loading would be placed
on it. We ran each of the simulations for a shorter duration
(approximately two hours) to ensure a value of cycles/CPU time
converged. This is then multiplied by $\mfom_C$ to return the FOM:
\begin{equation}
  \label{eq:fom_fomc}
  \mfom_C\cdot\frac{C}{t} =
  \frac{1}{\sigma(\hat{x})^2C}\cdot\frac{C}{t} =
  \frac{1}{\sigma(\hat{x})^2t} = \mfom\,.
\end{equation}
This process ensures that the FOM for all
runs is normalized to the same computer loading.

\subsection{Analysis Package}

As discussed in the previous section, we want to observe the
convergence behavior of the \gls{fom} to determine the final value. To
do so, we modified the Serpent 2 source code to generate
uniquely named output files at various cycle values. We developed the
\verb|WDT_Analysis| package\footnote{Available on GitHub
  \url{https://github.com/jsrehak/WDT_Analysis}} to leverage Python's
object oriented programming structure and enable easy analysis of the data. We utilize
a module, \verb|fom|, to analyze the convergence of \gls{fom} across
many Serpent 2 output files.

\subsection{Parameters of study}
Each Serpent 2 simulation generates hundreds of output parameters that
describe all processes that occur during particle propagation. To
focus our search, we identified two quantities that will be the focus
of this study: flux and total cross-section. Further study into other
parameters may reveal further benefit or issues with the \gls{wdt}
method. Both test cases have reflective boundary conditions, so we
will examine the infinite values for the parameters of interest. 

\subsubsection{PWR pin cell fast group}\label{sec:pwr_fast}

In the \gls{pwr} pin cell simulation, fast group interactions are
dominated by scattering and a small amount of absorption in the light
water moderator. The majorant cross-section is defined by the
uranium fuel; for values of $P_{\mathrm{wdt}} < 1.0$, only standard
delta-tracking is used in the fuel region. Therefore, any changes in
\gls{fom} are likely due to interactions in the moderator region.

\begin{figure}[hbtp]
      \begin{tabularx}{0.5\textwidth}{p{23px} p{165px}}
    & \centering      \large{PWR fast group $\phi_{\infty}$}
  \end{tabularx}
  \centering
      \includegraphics[width=0.45\textwidth]{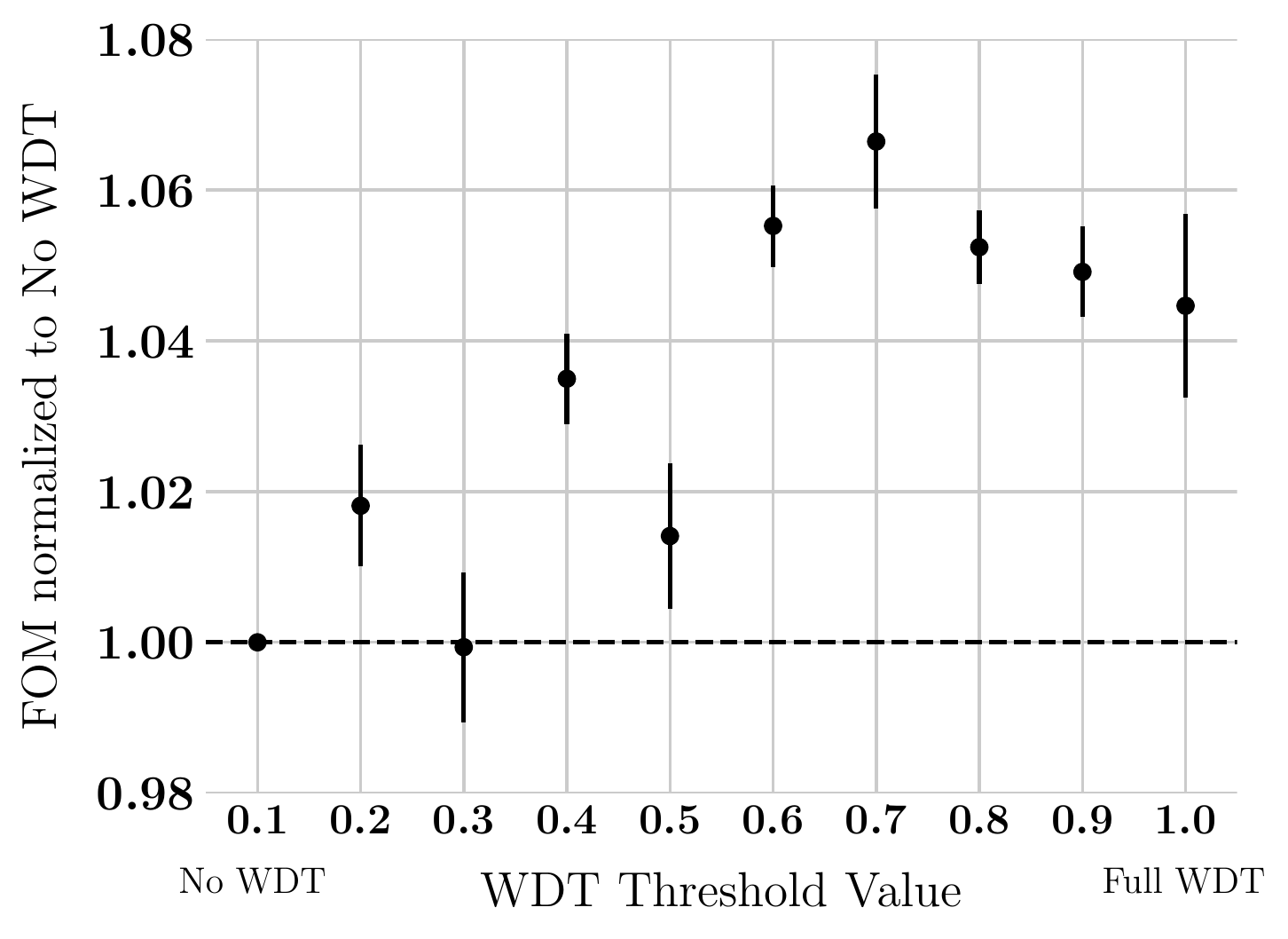}
      \caption{FOM for the PWR infinite flux $\phi_\infty$ for the
        fast group. Error bars shown are one standard
        deviation.}\label{fig:pwr_inf_flx_fast}
\end{figure}

The infinite flux ($\phi_\infty$) \gls{fom} of the fast group is shown
in Fig.~\ref{fig:pwr_inf_flx_fast}. The data show an improvement in
\gls{fom} at most values of $P_{\mathrm{wdt}}\leq 0.7$. As neutrons in
the fast group scatter in the moderator and lose energy, the total
cross-section for absorption increases exponentially leading to a
subsequent rise in $P_{\mathrm{real}}$. As discussed, the \gls{wdt}
routine is only used when $P_{\mathrm{real}}$ is less than or equal to
$P_{\mathrm{wdt}}$. Increasing $P_{\mathrm{wdt}}$ will therefore cause
more absorption events to use the \gls{wdt} routine instead of
standard delta-tracking. These absorption events now always contribute
to the statistics of the problem, reducing variance. This correlates
to the rise in average real collisions per track in
Fig.~\ref{fig:pwr_real_collisions}. Eventually, this increase in the
number of real collisions levels off for $P_{\mathrm{wdt}} > 0.7$, and
we see a subsequent leveling off of \gls{fom}. There is a
small decrease in \gls{fom}, possibly due to the inefficiency of
using \gls{wdt} compared to delta-tracking in high scattering regions.

\begin{figure}[hbtp]
    \begin{tabularx}{0.5\textwidth}{p{23px} p{165px}}
    & \centering\large{PWR fast group $\Sigma_{t,\infty}$}
  \end{tabularx}
  \centering
        
  \includegraphics[width=0.45\textwidth]{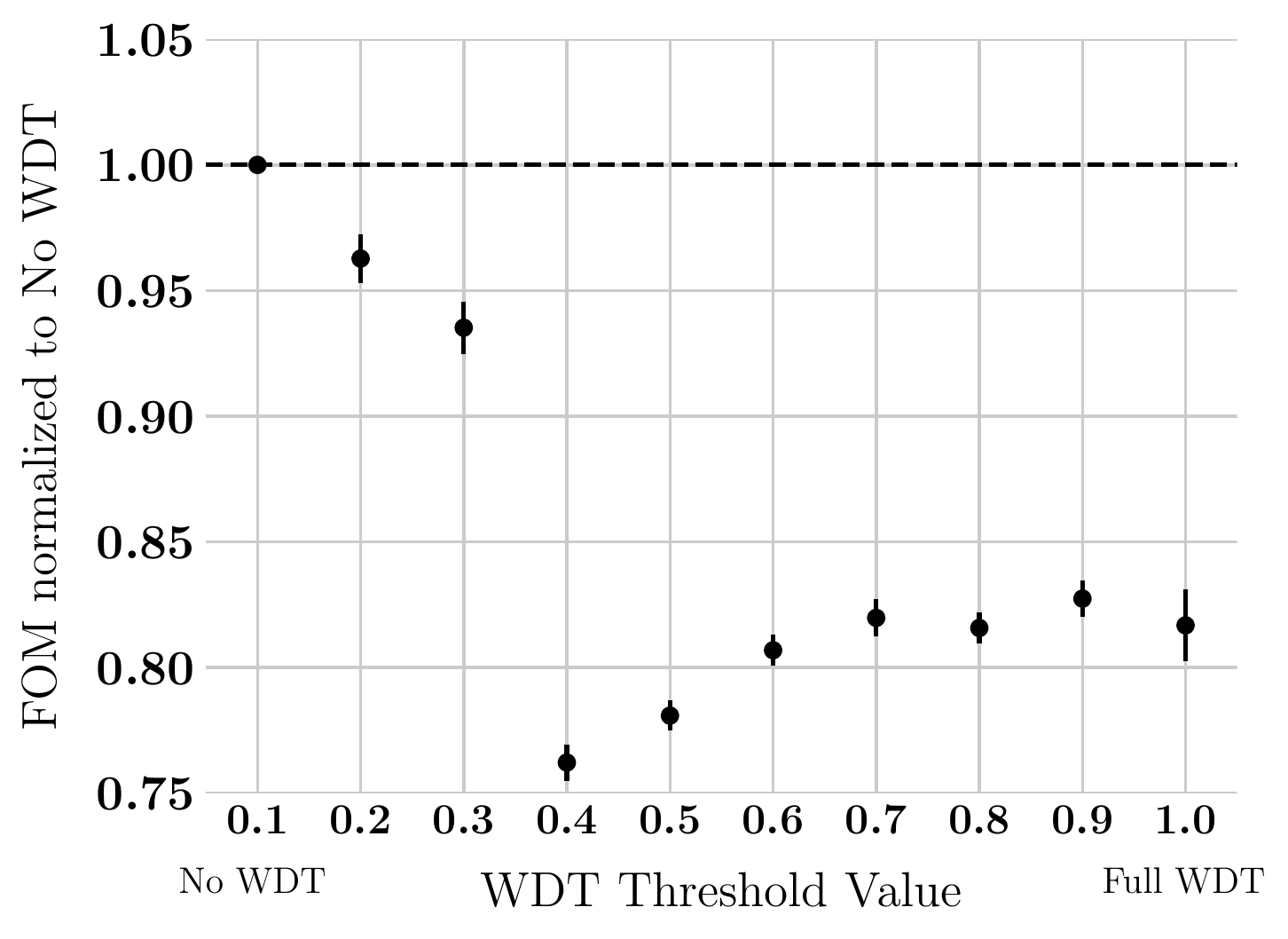}
  \caption{FOM for the PWR infinite total cross-section
    $\Sigma_{t,\infty}$ for the fast group. Error bars shown are one standard deviation.}\label{fig:pwr_inf_tot}
\end{figure}

The infinite total cross-section ($\Sigma_{\mathrm{tot, \infty}}$)
\gls{fom} of the fast group is shown in
Fig.~\ref{fig:pwr_inf_tot}. The reduction in \gls{fom} from the base
case mimics the pattern of the cycles per CPU time seen in
Fig.~\ref{fig:pwr_cyc_cpu}. This may occur because scattering is the
dominant contribution to the total cross-section. Thus, the collection
of more statistics for absorption events is overshadowed by the
inefficiencies in scattering. By not improving the variance
significantly, the driving factor in \gls{fom} becomes the cycles per
CPU time.

\subsubsection{PWR pin cell thermal group}\label{sec:pwr_thermal}

\begin{figure}[hbtp]
  \begin{tabularx}{0.5\textwidth}{p{23px} p{165px}}
    & \centering\large{PWR thermal group $\phi_{\infty}$}
  \end{tabularx}
  \centering
  \includegraphics[width=0.45\textwidth]{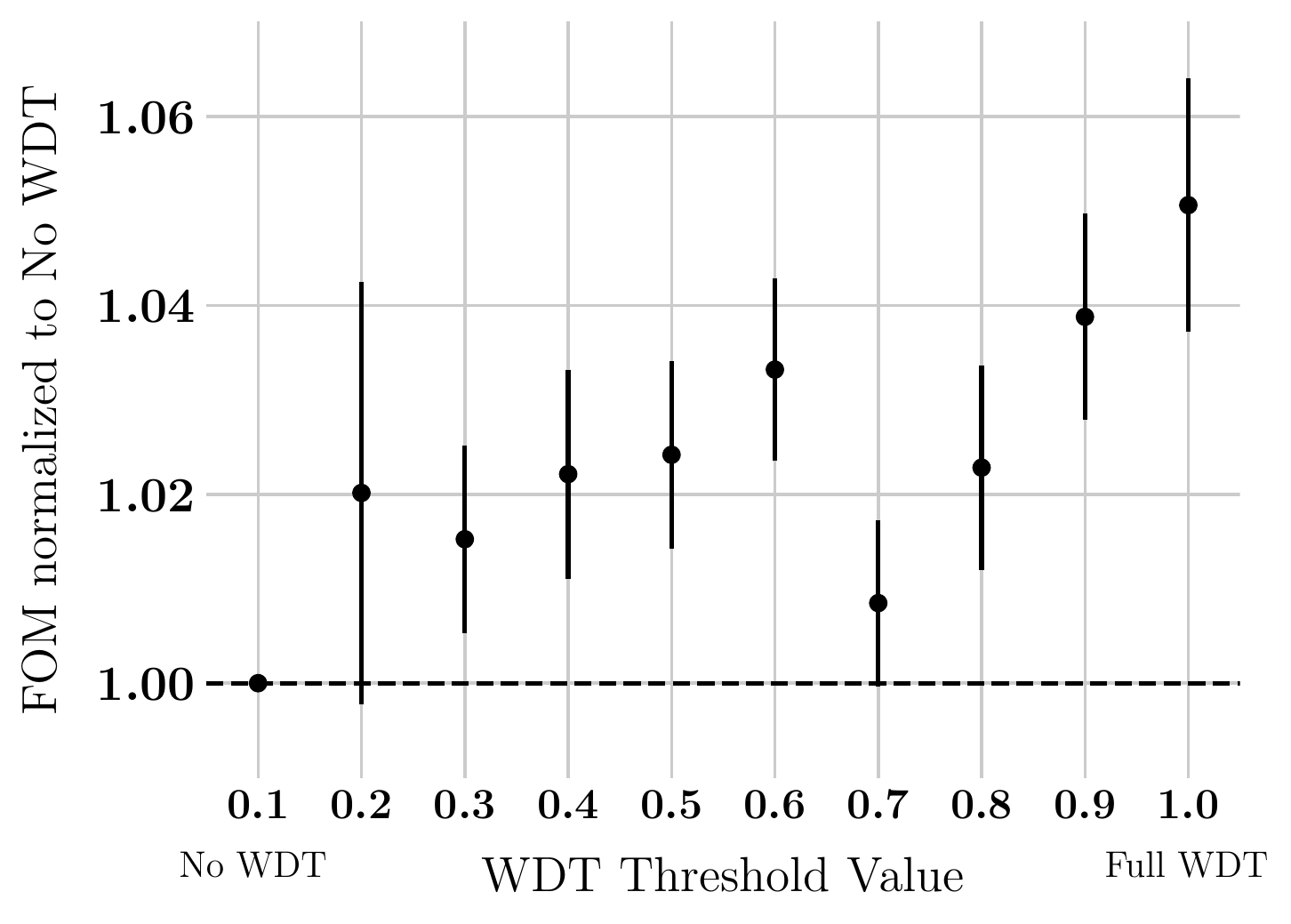}

  \caption{FOM for the PWR infinite flux $\phi_\infty$ for the
        thermal group. Error bars shown are one standard
        deviation.}\label{fig:pwr_inf_flx_thermal}
\end{figure}

The thermal group interactions in the \gls{pwr} pin cell simulation
are dominated by absorption, with very little scattering. For the
infinite flux, shown in Fig.~\ref{fig:pwr_inf_flx_thermal} we see a
consistent improvement in the \gls{fom} with increasing
$P_{\mathrm{wdt}}$. The improved statistics provided by the increased
number of real collisions outweighs the computational inefficiency of
the increased tallying required.

\begin{figure}[hbtp]
  \begin{tabularx}{0.5\textwidth}{p{23px} p{165px}}
    & \centering\large{PWR thermal group $\Sigma_{t,\infty}$}
  \end{tabularx}
  \centering
  \includegraphics[width=0.45\textwidth]{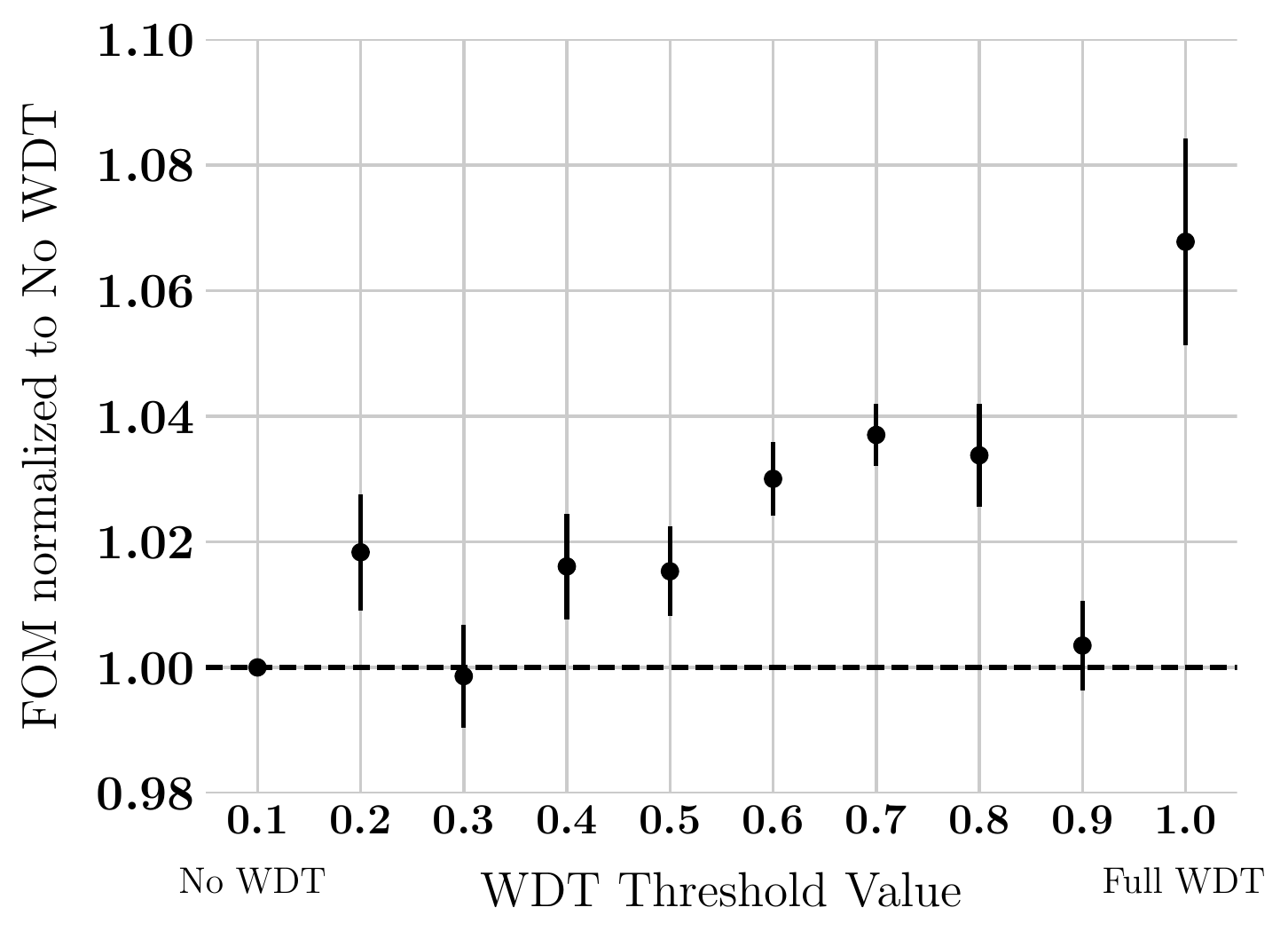}

  \caption{FOM for the PWR infinite total cross-section $\Sigma_{t,\infty}$ for the
        thermal group. Error bars shown are one standard
        deviation.}\label{fig:pwr_inf_tot_thermal}
\end{figure}

The infinite total cross-section for the PWR pin cell thermal
group is shown in Fig.~\ref{fig:pwr_inf_tot_thermal}. The trend
here is less consistent than in the other plots. There is a clear
increase in \gls{fom} at a $P_{\mathrm{wdt}}$ value of unity, the
only value where collisions in the fuel use \gls{wdt}. The fuel region
dominates the absorption events for thermal neutrons, so the improved
statistics leads to a higher \gls{fom}. Ignoring that point, we see a
rise and fall in \gls{fom} improvement, with a peak at
$P_{\mathrm{wdt}}=0.7$. Unlike the fast group, the total cross-section
is dominated by absorption, so the increased number of real absorption
events clearly improves the \gls{fom}. This improvement may not be
great enough to overcome the inefficiency introduced with more \gls{wdt}.
% wait, in this case the FOM is better for flux and xsec, so it is great enough...?
% is it that the fast case xsec FOM is so much worse?    
    
\subsubsection{Fast reactor pin cell fast group}
\label{sec:fast_pin_cell_fast_group}

In the fast reactor pin cell, the fast neutron group drives fission
and therefore dominates absorption reactions. Like the \gls{pwr} pin
cell, the majorant cross-section is defined by the MOX fuel. Changes
in \gls{fom} are therefore likely due to interactions in the lead
coolant and cladding. Lead has a very low absorption cross-section for
fast neutrons, so much of the \gls{fom} change will be driven by
absorption in the cladding.

\begin{figure}[hbtp]
  \begin{tabularx}{0.5\textwidth}{p{23px} p{165px}}
    & \centering\large{Fast reactor pin cell fast group $\phi_{\infty}$}
  \end{tabularx}
  \centering
  \includegraphics[width=0.45\textwidth]{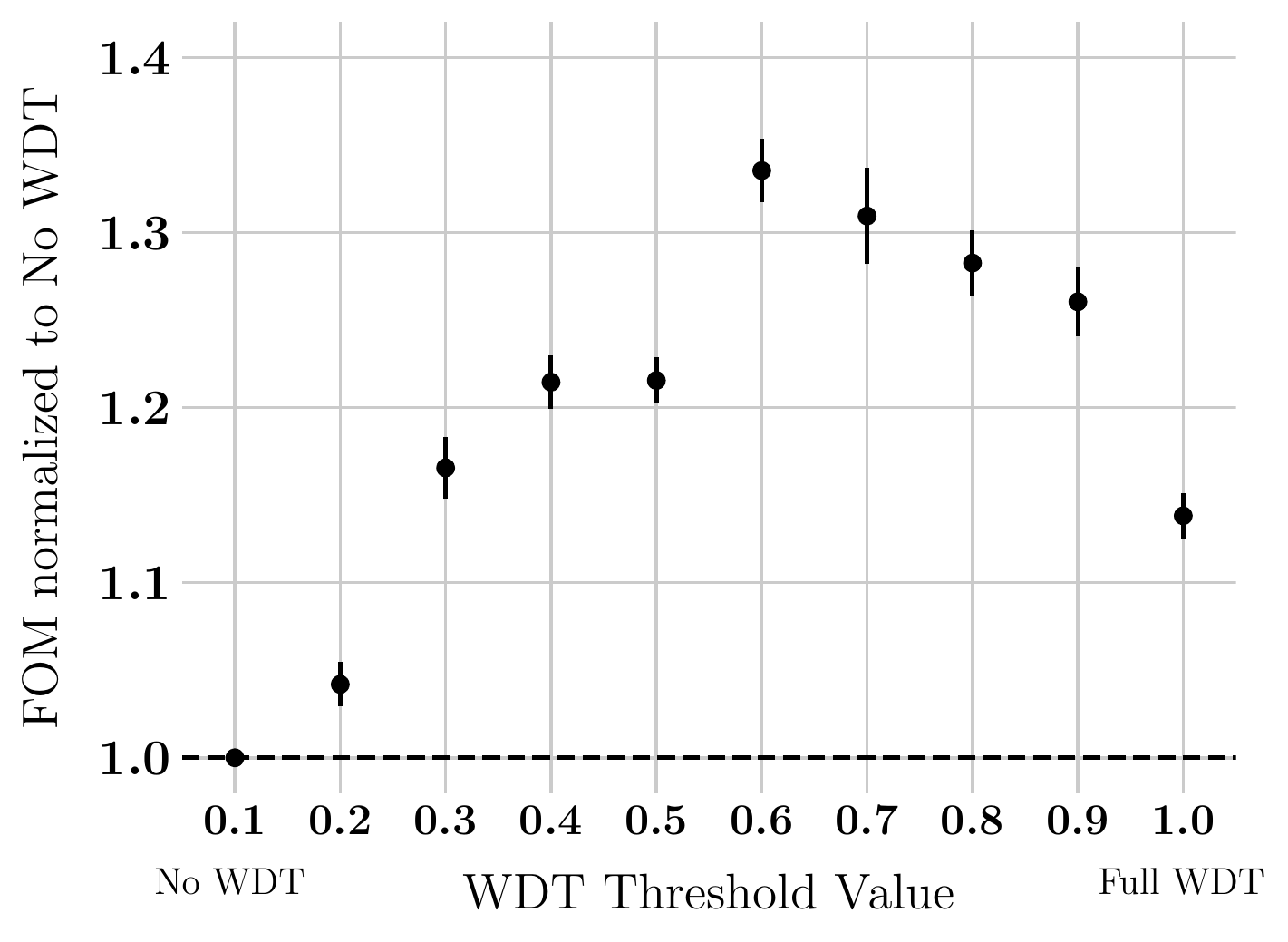}

  \caption{FOM for the fast reactor infinite flux
    $\phi_{\infty}$ for the fast group. Error bars shown are one
    standard deviation.}\label{fig:fast_inf_flx_fast}
\end{figure}

The infinite flux for the fast reactor pin cell is shown in
Fig.~\ref{fig:fast_inf_flx_fast}. We observe a rise in \gls{fom} as
more \gls{wdt} is introduced and more absorption events contribute to
the overall statistics. Eventually, the improvement in statistics is
overcome by the inefficiency of introducing many more real collisions,
and the improvement decreases. When \gls{wdt} is introduced in the
fuel, at $P_{\mathrm{wdt}}$ of unity we see a large decrease in
\gls{fom}, which matches the drop in cycles/CPU time at the same
value. 

\begin{figure}[hbtp]
  \begin{tabularx}{0.5\textwidth}{p{23px} p{165px}}
    & \centering\large{Fast reactor pin cell fast group $\Sigma_{t}$}
  \end{tabularx}
  \centering
  \includegraphics[width=0.45\textwidth]{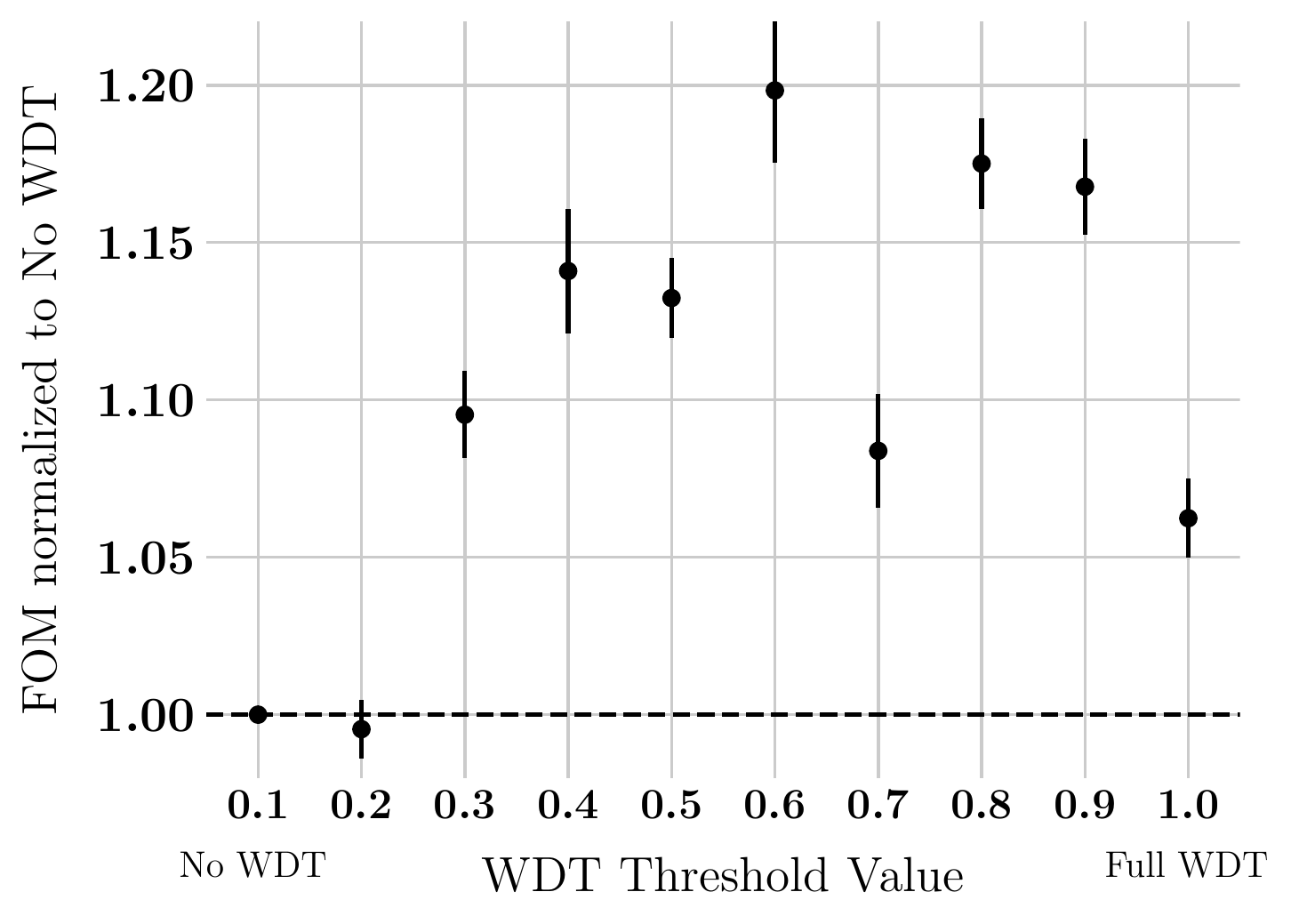}
  \caption{FOM for the fast reactor infinite total cross-section
    $\Sigma_{t}$ for the fast group. Error bars shown are one
    standard deviation.}\label{fig:fast_inf_tot_fast}
\end{figure}

A similar pattern is seen in the infinite total cross-section, shown
in Fig.~\ref{fig:fast_inf_tot_fast}. Again, we observe a rise in FOM
concurrent with the introduction of \gls{wdt}, as more absorption
events are contributing to the statistics. We see an uncharacteristic
drop at $P_{\mathrm{wdt}} = 0.7$ that does not seem to fit the overall
pattern. Further simulation and exploration is required to determine
the cause of this outlier.

\subsection{Choice of \gls{wdt} threshold}

As we see in the two test cases, use of the \gls{wdt} with scattering
routine results in an observable change in \gls{fom} in the infinite
flux and infinite total cross-section. Although the cycles per CPU
time is negatively impacted by \gls{wdt} for both test cases, the
collection of more statistics results in more efficient simulations,
as measured by \gls{fom}.

For the \gls{pwr} simulation, \gls{wdt} results in a small improvement
in the \gls{fom} for both group infinite fluxes, and the thermal group
infinite total cross-section. Therefore, for simulations seeking
accurate flux and thermal cross-sections, the data suggest that full
\gls{wdt} with scattering ($P_{\mathrm{wdt}} = 1$) would result in the
largest improvement. This improvement is driven by the use of the
\gls{wdt} routine in the fuel region, where \gls{wdt} only turns on when $P_{\mathrm{wdt}} = 1$ 
in our test cases. 
% I think that's what this means?
In simulations with heavy absorbers where the
fuel region does not define the majorant cross-section, the same
results may not be observed. In those cases, a lower value of
\gls{wdt} threshold will be sufficient to enable the \gls{wdt} routine
in the fuel regions.

Use of \gls{wdt} with scattering also generates an observed increase
in \gls{fom} for the fast reactor pin cell. The results differ from
the \gls{pwr} in two significant ways. First, the improvement is not
maximized at a threshold value of unity. Instead, improvement in
\gls{fom} for the infinite flux and total cross-section peaks when
$P_{\mathrm{wdt}} = 0.6$ and then declines. Second, the improvement is
of a much higher magnitude, reaching an improvement of nearly 20\% and
over 30\% for the infinite cross-section and flux, respectively. This
is much more significant than the sub-ten percent improvement for the
\gls{pwr} pin cell. Therefore, the data suggest that for a fast pin
cell simulation, mid-range value of \gls{wdt} threshold may result in
the greatest improvement in \gls{fom}.

A summary of all the relative \gls{fom} values and the associated
error are shown in Table~\ref{tab:fom}.

\begin{figure*}[hbtp]
  \caption{\gls{fom} as a function of $P_{\text{wdt}}$ for the PWR and fast
    reactor pin cell, noramlized to the no-WDT case of $P_{\text{wdt}}=0.1$.}\label{tab:fom}
  \centering
  \begin{tabular}{rrrrrrr}
    \toprule
    & \multicolumn{4}{c}{PWR} &
    \multicolumn{2}{c}{Fast Reactor}\\
    \cmidrule(lr){2-5} \cmidrule(lr){6-7}
    & \multicolumn{2}{c}{Fast group}
    & \multicolumn{2}{c}{Thermal group}
    & \multicolumn{2}{c}{Fast group} \\
    \cmidrule(lr){2-3} \cmidrule(lr){4-5} \cmidrule(lr){6-7}
    $P_\text{wdt}$ & $\phi_{\infty}$ & $\Sigma_{t,\infty}$ &
                   $\phi_{\infty}$  & $\Sigma_{t,\infty}$ & $\phi_{\infty}$  & $\Sigma_{t,\infty}$\\
    \midrule
    0.1 & 1.000 $\pm$  0.000  &  1.000 $\pm$  0.000&  1.000 $\pm$  0.000 &  1.000 $\pm$  0.000 &  1.000 $\pm$  0.000 &  1.000 $\pm$  0.000\\
    0.2 & 1.018 $\pm$  0.008  &  0.963 $\pm$  0.010&  1.020 $\pm$  0.022 &  1.018 $\pm$  0.009 &  1.042 $\pm$  0.013 &  0.995 $\pm$  0.009\\
    0.3 & 0.999 $\pm$  0.010  &  0.935 $\pm$  0.010&  1.015 $\pm$  0.010 &  0.999 $\pm$  0.008 &  1.166 $\pm$  0.018 &  1.095 $\pm$  0.014\\
    0.4 & 1.035 $\pm$  0.006  &  0.762 $\pm$  0.007&  1.022 $\pm$  0.011 &  1.016 $\pm$  0.008 &  1.217 $\pm$  0.015 &  1.140 $\pm$  0.019\\
    0.5 & 1.014 $\pm$  0.010  &  0.781 $\pm$  0.006&  1.024 $\pm$  0.010 &  1.015 $\pm$  0.007 &  1.216 $\pm$  0.013 &  1.132 $\pm$  0.013\\
    0.6 & 1.055 $\pm$  0.005  &  0.807 $\pm$  0.006&  1.033 $\pm$  0.010 &  1.030 $\pm$  0.006 &  1.335 $\pm$  0.018 &  1.198 $\pm$  0.023\\
    0.7 & 1.066 $\pm$  0.009  &  0.820 $\pm$  0.007&  1.008 $\pm$  0.009 &  1.037 $\pm$  0.005 &  1.310 $\pm$  0.028 &  1.084 $\pm$  0.018\\
    0.8 & 1.052 $\pm$  0.005  &  0.816 $\pm$  0.006&  1.023 $\pm$  0.011 &  1.034 $\pm$  0.008 &  1.283 $\pm$  0.019 &  1.175 $\pm$  0.014\\
    0.9 & 1.049 $\pm$  0.006  &  0.827 $\pm$  0.007&  1.039 $\pm$  0.011 &  1.003 $\pm$  0.007 &  1.261 $\pm$  0.020 &  1.168 $\pm$  0.015\\
    1.0 & 1.045 $\pm$  0.012  &  0.817 $\pm$  0.014&  1.051 $\pm$  0.013 &  1.068 $\pm$  0.017 &  1.138 $\pm$  0.013 &  1.062 $\pm$  0.013\\
    \bottomrule
\end{tabular}
\end{figure*}

\section{Conclusion}
\label{sec:conclusion}

In this work, we described the delta-tracking technique and the more
recent \gls{wdt} method that seeks to improve it. We then described
the issues with extending the \gls{wdt} to include scattering, and
introduced a novel hybrid modification. We implemented the method in the
Serpent 2 Monte Carlo code and examined the results of simulations
with two test cases. We then compared the \gls{fom} for two simulation
results against the base case to determine the impact of varying the
threshold at which \gls{wdt} with scattering is used.

Based on these results, we can conclude that the \gls{wdt} routine
with scattering does modestly improve \gls{fom} in infinite flux for
the \gls{pwr} and fast reactor pin cells. For total cross-section,
\gls{fom} is similarly improved, except in the fast group for the
\gls{pwr} pin cell. In both cases, the \gls{wdt} with scattering
routine reduces cycles per CPU time, by increasing the number of
simulated real collisions for absorption events. We hypothesize that
this increased number of real absorption events is the main driver in
improving \gls{fom}.

Further study should look into possible synergy or issues with use of
the \gls{wdt} with scattering routine and other variance reduction
techniques, such as implicit capture.

Based on the findings of Legrady \textit{et al.}\cite{Legrady2017}, we
expect an improved hybrid method could be developed that combines
sampling cross-sections not explored here. Extension of their method
in a highly scattering region may result in a similar multiplication
of particles, with a different splitting of weights. A hybrid method that uses
their optimized sampling cross-section for non-scattering events, and
normal delta-tracking for scattering events may result in significant
improvement over the method described here.

In addition, a finer resolution in threshold
values may provide a clearer relationship between the amount of
\gls{wdt} and improvement in \gls{fom}. Finally, we hypothesize that
the impact of \gls{wdt} on \gls{fom} will be very different in
simulations where the fuel region does not define the majorant
cross-section. Further study should look at test problems with heavy
absorbers that drive majorant cross-section.

\section{Acknowledgments}
This article was prepared by J.S. Rehak under award
NRC-HQ-84-14-G-0042 from the Nuclear Regulatory Commission. The
statements, finding, conclusions, and recommendations are those of the
author(s) and do not necessarily reflect the view of the US Nuclear
Regulatory Commission.

\section{References}
\label{sec:references}
\bibliographystyle{plain}
\bibliography{bib/bib}

\end{document}

%% file: gls/gls.tex
\newacronym{bwr}{BWR}{boiling water reactor}
\newacronym{cdf}{CDF}{cumulative distribution function}
\newacronym{cfe}{CFE}{collision flux estimator}
\newacronym{fom}{FOM}{figure of merit}
\newacronym{inl}{INL}{Idaho National Lab}
\newacronym{mox}{MOX}{Mixed Oxide}
\newacronym{pdf}{PDF}{probability density function}
\newacronym{pwr}{PWR}{pressurized water reactor}
\newacronym{tle}{TLE}{track-length estimator}
\newacronym{treat}{TREAT}{Transient Reactor Test Facility}
\newacronym{triso}{TRISO}{Tristructural-isotropic}
\newacronym{ucb}{Berkeley}{UC Berkeley}
\newacronym{vtt}{VTT}{VTT Technical Research Centre of Finland}
\newacronym{wdt}{WDT}{weighted delta-tracking}